# The Weight Distributions of Cyclic Codes and Elliptic Curves

Baocheng Wang, Chunming Tang, Yanfeng Qi, Yixian Yang, Maozhi Xu

*Abstract*—Cyclic codes with two zeros and their dual codes as a practically and theoretically interesting class of linear codes, have been studied for many years. However, the weight distributions of cyclic codes are difficult to determine. From elliptic curves, this paper determines the weight distributions of dual codes of cyclic codes with two zeros for a few more cases.

*Index Terms*—Cyclic codes, weight distribution, elliptic curves, Gaussian periods, linear codes.

## I. Introduction

Through this paper, let $p$ be a prime. Consider two positive integers $s$ and $m$, let $q = p^s$ and $r = q^m$. A $[n,k,d]$-linear code $\mathcal{C}$ over $GF(q)$ is a k-dimensional subspace of $GF(q)^n$ with minimum distance $d$. The linear code $\mathcal{C}$ is called a cyclic code if $\mathcal{C}$ is a cyclic set, that is, if $(c_0, c_1, \cdots, c_{n-1}) \in \mathcal{C}$, then $(c_{n-1}, c_0, \cdots, c_{n-2}) \in \mathcal{C}$. Consider the following correspondence:

$$\pi : GF(q)^n \longrightarrow GF(q)[x]/(x^n - 1)$$
$$(c_0, c_1, \cdots, c_{n-1}) \longmapsto c_0 + c_1 x + \cdots + c_{n-1} x^{n-1}.$$

Then we can identify a codeword $(c_0, c_1, \cdots, c_{n-1}) \in \mathcal{C}$ with the polynomial $c_0 + c_1 x + \cdots + c_{n-1} x^{n-1} \in GF(q)[x]/(x^n - 1)$. Note that $\pi(\mathcal{C})$ is a subset of $GF(q)[x]/(x^n-1)$. Then $\mathcal{C}$ is a cyclic code if and only if $\pi(\mathcal{C})$ is an ideal of $GF(q)[x]/(x^n-1)$. Since $GF(q)[x]/(x^n-1)$ is a principal ideal ring, then there exists a unique monic polynomial of the least degree $g(x)$ satisfying $\pi(\mathcal{C}) = <g(x)>$ and $g(x)|(x^n - 1)$. The polynomial $g(x)$ is called the generator polynomial of $\mathcal{C}$ and $h(x) = \frac{x^n-1}{g(x)}$ is called the parity-check polynomial of $\mathcal{C}$.

Let $A_i$ be the number of codewords with the Hamming weight $i$ in $\mathcal{C}$. The Hamming enumerator of $\mathcal{C}$ is the polynomial

$$A_0 + A_1 x + \cdots + A_x x^n.$$

And $(A_0, A_1, \cdots, A_n)$ is called the weight distribution of $\mathcal{C}$. Usually, it is difficult to determine the weight distribution of a cyclic code.

Let $\alpha$ be a generator of $GF(r)^*$. Consider a positive factor h of $(q-1)$ and a factor $e$ of $h$. Let $g = \alpha^{(q-1)/h}$, $\beta = \alpha^{(r-1)/e}$ and $n = h(r-1)/(q-1)$. Then the order of $g$ and $g^{-1}$ is $n$ and $(g\beta)^n = ((g\beta)^{-1})^n$. The minimal polynomials $m_{g^{-1}}(x)$ and $m_{(\beta g)^{-1}}(x)$ of $g^{-1}$ and $(\beta g)^{-1}$ are factors of $x^n - 1$. Since $e|h$, then

$$\frac{h}{e}\frac{r-1}{q-1} = \frac{h}{e}(q^{m-1} + q^{m-2} + \cdots + q + 1) > q^i - 1,$$

where $1 \leq i \leq m-1$. Hence, $g^{-1}$ and $(\beta g)^{-1}$ are not conjugates of each other. Then $m_{g^{-1}}(x)m_{(\beta g)^{-1}}(x)|(x^n-1)$. From Delsart's Theorem [7], we can give the trace representation of the cyclic code $\mathcal{C}_{(q,m,h,e)}$ with the parity-check polynomial $m_{g^{-1}}(x)m_{(\beta g)^{-1}}(x)$. Define

$$\mathbf{c}(a,b) = (Tr_{r/q}(ag^0 + b(\beta g)^0), Tr_{r/q}(ag^1 + b(\beta g)^1),$$
$$\cdots, Tr_{r/q}(ag^{n-1} + b(\beta g)^{n-1})),$$

where $Tr_{r/q}$ is the trace function from $GF(r)$ to $GF(q)$. Then we have

$$\mathcal{C}_{(q,m,h,e)} = \{\mathbf{c}(a,b) : (a,b) \in GF(r)\}.$$

The dimension of $\mathcal{C}_{(q,m,h,e)}$ is a factor of $2m$. Generally, the weight distribution of $\mathcal{C}_{(q,m,h,e)}$ is very complicated.

When $h = q-1$, the code $\mathcal{C}_{(q,m,h,e)}$ is the dual code of a primitive cyclic linear code with two zeros [3], [4], [5], [6], [15], [16], [18], [21]. The known results on the weight distribution of $\mathcal{C}_{(q,m,h,e)}$ are listed as follows:
1) $e > 1$ and $gcd(m, e(q-1)/h) = 1$ [14].
2) $e = 2$ and $gcd(m, e(q-1)/h) = 2$ [14].
3) $e = 2$ and $gcd(m, e(q-1)/h) = 3$ [8].
4) $e = 2$ and $p^j + 1 \equiv 0 \pmod{gcd(m, e(q-1)/h)}$, where $j$ is a positive integer [8].

When $e = 2$, Gaussian periods and cyclotomic numbers can be utilized to compute the weights and the corresponding frequencies, which determine the weight distribution of $\mathcal{C}_{(q,m,h,e)}$. When $e > 2$, it seems impossible to compute frequencies from cyclotomic numbers. In this paper, we use Gaussian periods and a class of elliptic curves to consider the weight distribution of the cyclic code $\mathcal{C}_{(q,m,h,e)}$ with $e = 3$.

## II. Cyclotomy, Gaussian periods and elliptic curves

Let $N$ be a positive factor of $r-1$ and $\alpha$ be a primitive element of $GF(r)$. Then the cyclotomic classes of order $N$ in $GF(r)$ are cosets

$$C_i^{(N,r)} = \alpha^i <\alpha^N>,$$

where $i$ is an integer. When $i \equiv j \pmod{N}$, $C_i^{(N,r)} = C_j^{(N,r)}$. The Gauss periods are

$$\eta_i^{(N,r)} = \sum_{x \in C_i^{(N,r)}} \chi(x),$$

B. Wang and Y, Yang are with Information Security Center, Beijing University of Posts and Telecommunications and Research Center on fictitious Economy and Data Science, Chinese Academy of Sciences, Beijing, 100088, China

C. Tang, Y. Qi and M. Xu is with the School of Mathematical Sciences , Peking University, 100871, China

C. Tang's e-mail: tangchunmingmath@163.com

where $\chi(x) = exp(2\pi\sqrt{-1}Tr_{r/p}(x))$, and $Tr_{r/p}$ is the trace function from $GF(r)$ to $GF(p)$.

Generally, it is difficult to compute the value of Gaussian periods. For some cases, the Gaussian periods can be computed. We list the following lemma on the Gaussian periods [17].

*Lemma 2.1:* When $N = 2$, the Gaussian periods are given by

$$\eta_0^{(2,r)} = \begin{cases} \frac{-1+(-1)^{sm-1}r^{1/2}}{2} & \text{if } p \equiv 1 \pmod 4 \\ \frac{-1+(-1)^{sm-1}(\sqrt{-1})^{sm}r^{1/2}}{2} & \text{if } p \equiv 3 \pmod 4 \end{cases}$$

and $\eta_1^{(2,r)} = -1 - \eta_0^{(2,r)}$.

For determining the weight distribution of $\mathcal{C}_{(q,m,h,e)}$, we introduce a class of elliptic curves [11].

**Definition** A twisted Jacobi intersection over the finite field $GF(r)$ is an elliptic curve defined by

$$J_{a,b} : \begin{cases} aU^2 + V^2 = T^2 \\ bU^2 + W^2 = T^2 \end{cases}$$

where $a, b$ are in $GF(r)$ satisfying $ab(a-b) \neq 0$.

Let $J_{a,b}(GF(r))$ be the set of all the $GF(r)$-rational points on $J_{a,b}$. For abbreviation, we often use the affine equation of $J_{a,b}$, that is,

$$\begin{cases} au^2 + v^2 = 1 \\ bu^2 + w^2 = 1 \end{cases}$$

On $J_{a,b}$, we have the following lemma [11].

*Lemma 2.2:* The twisted Jacobi intersection $J_{a,b}$ in affine coordinates over $GF(r)$

$$\begin{cases} au^2 + v^2 = 1 \\ bu^2 + w^2 = 1 \end{cases}$$

is birational equivalent to an elliptic curve in Weierstrass form

$$E_{a,b} : y^2 = x(x-a)(x-b).$$

This birational equivalence can be given by

$$\begin{cases} x = -\frac{a(w+1)}{v-1} \\ y = \frac{au}{v-1}(x-b) \end{cases}$$

and

$$\begin{cases} u = -\frac{2y}{x^2-ab} \\ v = \frac{x^2-2ax+ab}{x^2-ab} \\ w = \frac{x^2-2bx+ab}{x^2-ab} \end{cases}$$

Let $E$ be an elliptic curve defined over the finite field $GF(r)$. We have the following lemma on twisted curves of $E$ [10], [19].

*Lemma 2.3:* Let $E$ be an elliptic curve over $GF(r)$ defined by

$$E : y^2 = x^3 + a_2x^2 + a_4x + a_6.$$

Let $E'$ be another elliptic curve defined by

$$E' : y^2 = x^3 + \gamma a_2 x^2 + \gamma^2 a_4 x + \gamma^3 a_6,$$

where $\gamma$ is a quadratic nonresidue. Then $\#E(GF(r)) + \#E'(GF(r)) = 2(r+1)$.

In Lemma 2.3, $E'$ is called a quadratic twist of $E$.

## III. THE WEIGHT DISTRIBUTIONS FOR A CLASS OF CYCLIC CODES

In this section, we will determine the weight distributions for a class of cyclic codes. We first recall some notations defined above. $q = p^s$ and $r = q^m$, where $s$ and $m$ are two positive integers. $GF(r)^* = <\alpha>$. $h$ is a positive factor of $q-1$ and $e$ is a positive factor of $h$. $g = \alpha^{(q-1)/h}$, $\beta = \alpha^{(q-1)/e}$ and $n = h(r-1)/(q-1)$. $m_{g^{-1}}(x)$ and $m_{(\beta g)^{-1}}(x)$ are the minimum polynomials over $GF(q)$ of $g^{-1}$ and $(\beta g)^{-1}$. Let

$$\mathbf{c}(a,b) = (Tr_{r/q}(ag^0 + b(\beta g)^0), Tr_{r/q}(ag^1 + b(\beta g)^1),$$
$$\cdots, Tr_{r/q}(ag^{n-1} + b(\beta g)^{n-1})).$$

Then the cyclic code with the parity-check polynomial $m_{g^{-1}}(x)m_{(\beta g)^{-1}}(x)$ is

$$\mathcal{C}_{(q,m,h,e)} = \{\mathbf{c}(a,b) : a,b \in GF(r)\}.$$

For any $a, b \in GF(r)$, the Hamming weight of $\mathbf{c}(a,b)$ is $n - Z(r,a,b)$, where $Z(r,a,b) = \#\{x \in C_0^{((q-1)/h,r)} : Tr_{r/q}(f_{(a,b)}(x)) = 0\}$ and $f_{(a,b)}(x) = ax + \beta^{log_g(x)}bx$. From [14], [8], we have the following formula of $Z(r,a,b)$.

$$Z(r,a,b) = \frac{h(r-1)}{q(q-1)} + \frac{h}{eq}gcd(m,e(q-1)/h)\sum_{i=0}^{e-1}$$
$$\sum_{z \in C_{(q-1)i/h}^{(gcd(m,e(q-1)/h),r)}} \chi((a+\beta^i b)z).$$

When $e = 3$ and $gcd(m, e(q-1)/h) = 2$, we have the weight distribution of $\mathcal{C}_{(q,m,h,e)}$ in the following theorem.

*Theorem 3.1:* Let $h$ be a positive factor of $q-1$, $e = 3$ and $h \equiv 0 \pmod e$. Let $gcd(m, e(q-1)/h) = 2$, then the cyclic code $\mathcal{C}_{(q,m,h,e)}$ over $GF(q)$ is an $[n, 2m]$ code with the distribution weight in Table I.

TABLE I
THE WEIGHT DISTRIBUTION FOR THE CASE $e = 3$ AND $gcd(m, e(q-1)/h) = 2$

| Weight | Frequency |
|---|---|
| 0 | 1 |
| $\frac{2h}{3}[q^{m-1} + q^{(m-2)/2}]$ | $\frac{3(r-1)}{2}$ |
| $\frac{2h}{3}[q^{m-1} - q^{(m-2)/2}]$ | $\frac{3(r-1)}{2}$ |
| $h[q^{m-1} + q^{(m-2)/2}]$ | $\frac{(r-1)(r-5)}{8}$ |
| $h[q^{m-1} - q^{(m-2)/2}]$ | $\frac{(r-1)(r-5)}{8}$ |
| $\frac{h}{3}[3q^{m-1} + q^{(m-2)/2}]$ | $\frac{3(r-1)^2}{8}$ |
| $\frac{h}{3}[3q^{m-1} - q^{(m-2)/2}]$ | $\frac{3(r-1)^2}{8}$ |

*Proof:* Let

$$Y(r,a,b) = \sum_{i=0}^{e-1}\sum_{z \in C_{(q-1)i/h}^{(gcd(m,e(q-1)/h),r)}} \chi((a+\beta^i b)z).$$

Since $gcd(m, e(q-1)/h) = 2$, then $\frac{q-1}{h} \equiv 0 \pmod 2$. Further, for any $i$, $\frac{(q-1)i}{h} \equiv 0 \pmod 2$. Then

$$Y(r,a,b) = \sum_{i=0}^{e-1}\sum_{z \in C_0^{(2,r)}} \chi((a+\beta^i b)z).$$

Then
$$Y(r,a,b) = \sum_{z \in C_0^{(2,r)}} \chi((a+b)z) + \sum_{z \in C_0^{(2,r)}} \chi((a+\beta b)z)$$
$$+ \sum_{z \in C_0^{(2,r)}} \chi((a+\beta^2 b)z).$$

Hence, to determine the weight distribution, we just need to determine the distribution of $Y(r,a,b)$.

Note that the values of $\sum_{z \in C_0^{(2,r)}} \chi((a+b)z)$, $\sum_{z \in C_0^{(2,r)}} \chi((a+\beta b)z)$ and $\sum_{z \in C_0^{(2,r)}} \chi((a+\beta^2 b)z)$ lie in the set $\{\eta_0^{(2,r)}, \eta_1^{(2,r)}, \frac{r-1}{2}\}$. If either $a$ or $b$ is not zero, then $a+b$, $a+\beta b$ and $a+\beta^2 b$ have at most one zero. For further discussion, we list some notations here.

$\mathcal{C}_0^* = \{\mathbf{c}(a,b) : a+b, a+\beta b$ and $a+\beta^2 b$ have just one zero and the orther two lie in $C_0^{(2,r)}$ $\}$.

$\mathcal{C}_2^* = \{\mathbf{c}(a,b) : a+b, a+\beta b$ and $a+\beta^2 b$ have just one zero and the orther two lie in $C_1^{(2,r)}$ $\}$.

$\mathcal{C}_0 = \{\mathbf{c}(a,b) :$ All of $a+b, a+\beta b$ and $a+\beta^2 b$ lie in $C_0^{(2,r)}$ $\}$.

$\mathcal{C}_1 = \{\mathbf{c}(a,b) :$ One of $a+b, a+\beta b$ and $a+\beta^2 b$ lies in $C_1^{(2,r)}$. The other two lie in $C_0^{(2,r)}$ $\}$.

$\mathcal{C}_2 = \{\mathbf{c}(a,b) :$ One of $a+b, a+\beta b$ and $a+\beta^2 b$ lies in $C_0^{(2,r)}$. The other two lie in $C_1^{(2,r)}$ $\}$.

$\mathcal{C}_3 = \{\mathbf{c}(a,b) :$ All of $a+b, a+\beta b$ and $a+\beta^2 b$ lie in $C_1^{(2,r)}$. $\}$.

Any two sets defined above do not intersect. If one of $a+b$, $a+\beta b$ and $a+\beta^2 b$ is zero, we assume that $a+b=0$. Then $a+\beta b = b(-1+\beta)$, $a+\beta^2 b = b(-1+\beta^2)$ and $\frac{a+\beta b}{a+\beta^2 b} = \frac{1}{1+\beta} = -\beta^2$. Note that all the elements in $GF(q)$ are quadratic residues and $-\beta^2$ is a quadratic residue. Hence, $a+\beta b$ and $a+\beta^2 b$ are both quadratic residues or both quadratic nonresidues. We can also discuss the case $a+\beta b = 0$ or $a+\beta^2 b = 0$. Hence, if one of $a+b$, $a+\beta b$ and $a+\beta^2 b$ is zero, the other two are both quadratic residues or both quadratic nonresidues. Precisely, these sets become a partition of $\mathcal{C}_{(q,m,h,e)} - \{\mathbf{c}(0,0)\}$.

We construct the following bijective maps for these sets.

$$\begin{aligned} \mathcal{C}_0^* &\longrightarrow \mathcal{C}_2^* \\ \mathbf{c}(a,b) &\longmapsto \mathbf{c}(\alpha a, \alpha b) \\ \mathcal{C}_0 &\longrightarrow \mathcal{C}_3 \\ \mathbf{c}(a,b) &\longmapsto \mathbf{c}(\alpha a, \alpha b) \\ \mathcal{C}_1 &\longrightarrow \mathcal{C}_2 \\ \mathbf{c}(a,b) &\longmapsto \mathbf{c}(\alpha a, \alpha b) \end{aligned}$$

Thus we have
$$\#\mathcal{C}_0^* = \#\mathcal{C}_2^*, \quad \#\mathcal{C}_0 = \#\mathcal{C}_3, \quad \#\mathcal{C}_1 = \#\mathcal{C}_2.$$

Now we first consider the set $\mathcal{C}_0^*$. Assume that $a+b=0$. Then $a+\beta b = b(-1+\beta)$ and $a+\beta^2 b = b(-1+\beta^2)$. $-1+\beta$ and $-1+\beta^2$ in $GF(q)$ are both quadratic residues. Hence, $\mathbf{c}(a,b) \in \mathcal{C}_0^*$ if and only if $b \in C_0^{(2,r)}$. The number of $\mathbf{c}(a,b)$ satisfying this condition is $\frac{r-1}{2}$. We can also discuss the case $a+\beta b = 0$ or $a+\beta^2 b$. Hence,

$$\#\mathcal{C}_0^* = \#\mathcal{C}_2^* = \frac{3(r-2)}{2}$$

.

Then we consider the set $\mathcal{C}_0$.

1) The number of codewords of the form $\mathbf{c}(a,0)$ in $\mathcal{C}_0$ is $\frac{r-1}{2}$.

2) When $b \neq 0$, we have
$$a+b = b(\frac{a}{b}+1), a+\beta b = b(\frac{a}{b}+\beta), a+\beta^2 b = b(\frac{a}{b}+\beta^2).$$

We introduce two auxiliary sets for counting $\mathcal{C}_0$.
$S_0 = \{c \in GF(r) : c+1, c+\beta, c+\beta^2 \in C_0^{(2,r)}\}$.
$S_3 = \{c \in GF(r) : c+1, c+\beta, c+\beta^2 \in C_1^{(2,r)}\}$.
The number of codewords satisfying $b \neq 0$ in $\mathcal{C}_0$ is
$$\frac{r-1}{2}(\#S_0 + \#S_3).$$

Hence, we have
$$\#\mathcal{C}_0 = \frac{r-1}{2}(\#S_0 + \#S_3 + 1).$$

To compute $\#S_0$, we introduce the following system of equations.
$$\begin{cases} c+1 = u^2 \\ c+\beta = v^2 \\ c+\beta^2 = w^2 \end{cases}$$

This system of equations is equivalent to the following system of equations.
$$\begin{cases} u^2 - v^2 = 1 - \beta \\ u^2 - w^2 = 1 - \beta^2 \end{cases} \quad (1)$$

Hence, we get
$$\#S_0 = \frac{1}{8}\#(\{(u,v,w) : u,v,w \in GF(r), (u,v,w)$$
$$\text{is the solution of } (1), uvw \neq 0\}).$$

To compute the number of the solution set of this system of equations, we consider the curve in projective coordinate corresponding to the system of equations (1).
$$\begin{cases} U^2 - V^2 = (1-\beta)Z^2 \\ U^2 - W^2 = (1-\beta^2)Z^2 \end{cases}$$

We denote this curve by $J_0$. The affine equation of $J_0$ can be converted into the following equations.
$$\begin{cases} (\frac{u}{\sqrt{1-\beta}})^2 + (\frac{v}{\sqrt{-(1-\beta)}})^2 = 1 \\ \frac{1}{1+\beta}(\frac{u}{\sqrt{1-\beta}})^2 + (\frac{w}{\sqrt{-(1-\beta^2)}}) = 1. \end{cases}$$

Note that $1+\beta+\beta^2 = 0$, $\beta \in GF(q)$ and all the elements in $GF(q)$ are quadratic residues. Hence, $J_0$ is birational equivalent over $GF(r)$ to the following curve
$$\begin{cases} u^2 + v^2 = 1 \\ -\beta u^2 + w^2 = 1 \end{cases}$$

From Lemma 2.2, $J_0$ is birational equivalent over $GF(r)$ to the elliptic curve
$$\begin{aligned} y^2 &= x(x-1)(x+\beta) \\ &= x^3 + (\beta-1)x^2 - \beta x \\ &= (x+\frac{\beta-1}{3})^3 + (\sqrt{-\frac{\beta-1}{3}})^6. \end{aligned}$$

Thus, $J_0$ is birational equivalent over $GF(r)$ to
$$y^2 = x^3 + 1.$$

Similarly, to compute $\#S_3$, we introduce the following system of equations.
$$\begin{cases} c+1 = \alpha u^2 \\ c+\beta = \alpha v^2 \\ c+\beta^2 = \alpha w^2 \end{cases}$$

This systems of equations is equivalent to the following systems of equations.
$$\begin{cases} \alpha u^2 - \alpha v^2 = 1 - \beta \\ \alpha u^2 - \alpha w^2 = 1 - \beta^2 \end{cases} \quad (2)$$

Hence, we get
$$\#S_3 = \frac{1}{8}\#(\{(u,v,w): u,v,w \in GF(r), (u,v,w)$$
$$\text{is the solution of } (2), uvw \neq 0\}).$$

The system of equations (2) stands for an elliptic curve, which is denoted by $J_3$. Similar to the discussion of $J_0$, $J_3$ is birational equivalent over $GF(r)$ to the following curve.
$$\begin{cases} \alpha u^2 + \alpha v^2 = 1 \\ -\alpha\beta u^2 + \alpha w^2 = 1 \end{cases}$$

This curve can be transformed into the following curve.
$$\begin{cases} 1 + (\frac{v}{u})^2 = \alpha(\frac{1}{\alpha u})^2 \\ -\beta + (\frac{w}{u})^2 = \alpha(\frac{1}{\alpha u})^2 \end{cases}$$

that is,
$$\begin{cases} \alpha(\frac{1}{\alpha u})^2 - (\frac{v}{u})^2 = 1 \\ -\frac{\alpha}{\beta}(\frac{1}{\alpha u})^2 + \frac{1}{\beta}(\frac{w}{u})^2 = 1 \end{cases}$$

Thus, $J_3$ is birational equivalent over $GF(r)$ to the following curve.
$$\begin{cases} \alpha u^2 + v^2 = 1 \\ -\alpha\beta^2 u^2 + w^2 = 1 \end{cases}$$

From Lemma 2.2, $J_0$ is birational equivalent over $GF(r)$ to the elliptic curve.
$$y^2 = x(x-\alpha)(x+\alpha\beta^2)$$
$$= x^3 + \alpha(\beta^2-1)x^2 - \alpha^2\beta^2 x$$
$$= (x + \frac{\alpha(\beta^2-1)}{3})^3 + (\sqrt{-\frac{(\beta^2-1)}{3}})^6 \alpha^3.$$

Hence, $J_3$ is birational equivalent over $GF(r)$ to the elliptic curve.
$$y^2 = x^3 + \alpha^3.$$

From Lemma 2.3, $y^2 = x^3 + \alpha^3$ is a quadratic twist of $y^2 = x^3 + 1$. And we have
$$\#J_0(GF(r)) + \#J_3(GF(r)) = 2(r+1). \quad (3)$$

The infinite points on $J_0$ satisfy
$$\begin{cases} U^2 - V^2 = (1-\beta)Z^2 = 0 \\ U^2 - W^2 = (1-\beta^2)Z^2 = 0 \end{cases}$$

Then, the curve $J_0$ has four infinite points $(U:V:W:Z) = (1:\pm 1:\pm 1:0)$. In the system of equations (1), if $u = 0$, then
$$v^2 = -1+\beta, \quad w^2 = -1+\beta^2.$$

Thus, (1) has four solutions $(u,v,w) = (0, \pm\sqrt{-1+\beta}, \pm\sqrt{-1+\beta^2})$ satisfying $u = 0$. Further, (1) has four solutions satisfying $v = 0$ and four solutions satisfying $w = 0$. Hence, we have
$$\#S_0 = \frac{1}{8}(\#J_0(GF(r)) - 16). \quad (4)$$

The curve $J_3$ also has four infinite points. The system of equations (2) does not have any solution $(u,v,w)$ satisfying $uvw = 0$. Hence, we have
$$\#S_3 = \frac{1}{8}(\#J_3(GF(r)) - 4). \quad (5)$$

Then from Equation (3), (4) and (5), we have
$$\#S_0 + \#S_3 = \frac{1}{8}(\#J_0(GF(r)) + \#J_3(GF(r)) - 20)$$
$$= \frac{1}{4}(r-9).$$

Hence, we obtain
$$\#\mathcal{C}_0 = \#\mathcal{C}_3 = \frac{r-1}{2}\frac{r-5}{4}.$$

If either $a$ and $b$ is not zero, $Z(r,a,b) < n$ and $\mathbf{c}(a,b) \neq 0$. Hence, $\mathcal{C}_{(q,m,h,e)}$ has the dimension $2m$ and it has $r^2 - 1$ nonzero codewords. Then we obtain
$$\#\mathcal{C}_1 = \#\mathcal{C}_2 = \frac{r^2 - 1 - \#\mathcal{C}_0^* - \#\mathcal{C}_2^* - \#\mathcal{C}_0 - \#\mathcal{C}_3}{2}$$
$$= \frac{3(r-1)^2}{8}.$$

From the above discussion, we have the distribution of $Y(r,a,b)$ in Table II.

TABLE II
THE DISTRIBUTION OF $Y(r,a,b)$

| Value | Frequency |
|---|---|
| $\frac{3(r-1)}{2}$ | 1 |
| $\frac{r-1}{2} + 2\eta_0^{(2,r)}$ | $\frac{3(r-1)}{2}$ |
| $\frac{r-1}{2} + 2\eta_1^{(2,r)}$ | $\frac{3(r-1)}{2}$ |
| $3\eta_0^{(2,r)}$ | $\frac{(r-1)(r-5)}{8}$ |
| $3\eta_1^{(2,r)}$ | $\frac{(r-1)(r-5)}{8}$ |
| $-1 + \eta_0^{(2,r)}$ | $\frac{3(r-1)^2}{8}$ |
| $-1 + \eta_1^{(2,r)}$ | $\frac{3(r-1)^2}{8}$ |

Note that the Hamming weight of $\mathbf{c}(a,b)$ is $n - Z(r,a,b)$. From Lemma 2.1, we can obtain the weight distribution in Table I. Hence, this theorem follows. ∎

**Example** Let $q = 7$, $m = 2$, $e = 3$ and $h = 3$. Then $gcd(m, e(q-1)/h) = 2$. From Theorem 3.1, the cyclic code $\mathcal{C}_{(q,m,h,e)}$ is an [24,4,12]-linear code over $GF(7)$ with the following weight distribution.

$$1 + 72x^{12} + 72x^{16} + 264x^{18} + 864x^{20} + 864x^{22} + 264x^{24}.$$

## IV. Conclusion

The weight distribution of $\mathcal{C}_{(q,m,h,e)}$ has been determined in the following cases:

1) $e > 1$ and $gcd(m, e(q-1)/h) = 1$ [14].
2) $e = 2$ and $gcd(m, e(q-1)/h) = 2$ [14].
3) $e = 2$ and $gcd(m, e(q-1)/h) = 3$ [8].
4) $e = 2$ and $p^j + 1 \equiv 0 \pmod{gcd(m, e(q-1)/h)}$, where $j$ is a positive integer [8].
5) $e = 3$ and $gcd(m, e(q-1)/h) = 2$ (This paper).

Although Gaussian periods of orders 3,4,5,6,8,12 [12], [13], [17] and Gaussian periods in the semi-primitive [1], [17] and quadratic-residue cases [2], [17] can be determined, it is still difficult to determine the weight distribution of $\mathcal{C}_{(q,m,h,e)}$. Let $gcd(m, e(q-1)/h) = f$. From our discussion, the frequency of the weight in $\mathcal{C}_{(q,m,h,e)}$ is related to the distribution of rational points on the following family of curves over $GF(r)$.

$$\begin{cases} x + \beta^0 &= \alpha_0 x_0^f \\ x + \beta^1 &= \alpha_1 x_1^f \\ x + \beta^2 &= \alpha_2 x_2^f \\ \cdots\cdots\cdots\cdots \\ x + \beta^{e-1} &= \alpha_{e-1} x_{e-1}^f \end{cases} \qquad (6)$$

where $\alpha_0, \alpha_1, \cdots, \alpha_{e-1} \in \{\alpha^0, \alpha^1, \cdots, \alpha^{e-1}\}$.

When $e = 1$, (6) is a system of linear equations, whose solution set can be easily determined. When $e = 2$, (6) is equivalent to the curve $\alpha_0 x_0^f - \alpha_1 x_1^f = \beta^0 - \beta^1$. To compute the number of solutions of the diagonal homogenous equation is equivalent to compute cyclotomic numbers [9], [20]. When $e = 3$ and $f = 2$, from our discussion we just consider elliptic curves in Jacobi form, that is, twisted Jacobi intersections. Generally, when $e \geq 3$ and $f \geq 2$, we should consider a class of curves of high degree over high dimensional spaces. It is difficult to find the distribution of rational points on these curves. Hence, generally, to determine the weight distributions of duals of cyclic codes with two zeros is difficult. The reader is invited to attack the weight distribution problem for the open cases.

## Acknowledgment

The authors would like to thank Prof. Cunsheng Ding for his help and advice in this work. Baocheng Wang and Yixian Yan acknowledge support from National Science Foundation of China Innovative Grant (70921061), the CAS/SAFEA International Partnership Program for Creative Research Teams. Chungming Tang, Yanfeng Qi and Maozhi Xu acknowledge support from the Natural Science Foundation of China (Grant No.10990011 & No.60763009).

## References

[1] L. D. Baumert, W. H. Mills and R. L. Ward, Uniform cyclotomy, J. Number Theory, vol. 14, pp. 67-82, 1982.
[2] L. D. Baumert and J. Mykkeltveit, Weight distributions of some irreducible cyclic codes, DSN Progress Report, vol. 16, pp. 128-131, 1973.
[3] N. Boston, G. McGuire, The weight distributions of cyclic codes with two zeros and zeta functions, Journal of Symbolic Computation, vol. 45, no. 7, pp. 723-733, 2010.
[4] A. Canteaut, P. Charpin, H. Dobbertin, Weight divisibility of cyclic codes, highly nonlinear functions on $F_{2^m}$; and crosscorrelation of maximum-length sequences, SIAM J. Discrete Math., vol. 13, pp. 105-138, 2000.
[5] C. Carlet, P. Charpin and V. Zinoviev, Codes, bent functions and permutations suitable for DES-like cryptosystems, Des Codes Crypt, vol. 15, pp. 125-156, 1998.
[6] P. Charpin, Cyclic codes with few weights and Niho exponents, J. Comb. Theory Ser. A, vol. 108, pp. 247-259, 2004.
[7] P. Delsarte, On subfield subcodes of modified Reed-Solomon codes, IEEE Trans. Inform. Theory, vol. 21, no. 5, pp. 575-576, Sep. 1975.
[8] C. Ding, Y. Liu, C. Ma, L. Zeng, The weight distributions of the duals of cyclic codes with two zeros, IEEE Trans. Inform. Theory, to appear.
[9] C. Ding and J. Yin, Sets of optimal frequency hopping sequences, IEEE Trans. Inform. Theory, vol. 54, no. 8, pp. 3741-3745, August 2008.
[10] A. Enge, Elliptic curves and their applications to cryptography: An introduc- tion. Boston, Dordrecht, London: Kluwer Acad. Pub. 1999
[11] R.Q. Feng, M.L. Nie, H.F Wu, Twisted Jacobi Intersections Curves, Theory and applications of models of computation, Lecture Notes in Computer Science, 2010, Volume 6108/2010, 199-210
[12] S. J. Gurak, Periodic polynomials for Fq of fixed small degree, CRM Proceedings and Lecture Notes, vol. 36, pp. 127-145, 2004.
[13] A. Hoshi, Explicit lifts of quintic Jacobi sums and periodic polynomials for Fq, Proc. Japan Acad., vol. 82, Ser. A, pp. 87-92, 2006.
[14] C. Ma, L. Zeng, Y. Liu, D. Feng and C. Ding, The weight enumerator of a class of cyclic codes, IEEE Trans. Inform. Theory, Vol. 57, No. 1, January 2011
[15] G. McGuire, On three weights in cyclic codes with two zeros, Finite Fields and Their Applications, vol. 10, pp. 97-104, 2004.
[16] M. Moisio and K. Ranto, Kloosterman sum identities and low-weight codewords in a cyclic code with two zeros, Finite Fields and Their Applications, vol. 13, pp. 922-935, 2007.
[17] G. Myerson, Period polynomials and Gauss sums for finite fields, Acta Arith., vol. 39, pp. 251-264, 1981.
[18] R. Schroof, Families of curves and weight distribution of codes, Bulletin of the American Mathematical Society, vol. 32, no. 2, pp. 171-183, 1995.
[19] J. H. Silverman, The arithmetic of elliptic curves. GTM 106. New York: Springer-Verlag, 1985
[20] T. Storer, Cyclotomy and Difference Sets, Marham, Chicago, 1967.
[21] J. Yuan, C. Carlet and C. Ding, The weight distribution of a class of linear codes from perfect nonlinear functions, IEEE Trans. Inform. Theory, vol. 52, no. 2, pp. 712-717, Feb. 2006